\documentclass{article}

\usepackage{arxiv}

\usepackage[utf8]{inputenc} 
\usepackage[T1]{fontenc}    
\usepackage{hyperref}       
\usepackage{url}            
\usepackage{booktabs}       
\usepackage{amsfonts}       
\usepackage{nicefrac}       
\usepackage{microtype}      
\usepackage{lipsum}		
\usepackage{graphicx}
\usepackage{natbib}
\usepackage{doi}
\usepackage{tabularx} 
\renewcommand{\arraystretch}{1.2} 
\usepackage{amsmath}
\title{A Gossip-Enhanced Communication Substrate for Agentic AI: Toward Decentralized Coordination in Large-Scale Multi-Agent Systems}

\author{ 
    Nafiul I. Khan \\
	Department of Computer Technology\\
	City Politechnique College\\
	Khulna, Bangladesh \\
	\texttt{earthkhan01@gmail.com} \\
	\AND
    \href{https://orcid.org/0000-0001-9051-1370}{\includegraphics[scale=0.06]{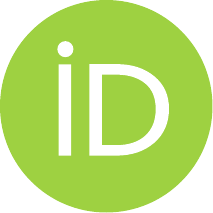}\hspace{1mm}Mansura Habiba} \\
	Principal Platform Architect - Agentic AI\\
	IBM Software\\
	Dublin, Ireland \\
	\texttt{mansura.habiba@gmail.com} \\
    \And
    Rafflesia Khan\\
	Software Developer\\
	IBM Software\\
	Dublin, Ireland \\
	\texttt{khan.rafflesia@gmail.com} \\
}

\renewcommand{\headeright}{}
\renewcommand{\shorttitle}{Gossip as Communication Fabric}

\hypersetup{
pdftitle={A template for the arxiv style},
pdfsubject={q-bio.NC, q-bio.QM},
pdfauthor={David S.~Hippocampus, Elias D.~Striatum},
pdfkeywords={First keyword, Second keyword, More},
}

\begin{document}
\maketitle

\begin{abstract}
As agentic platforms scale, agents are moving beyond fixed roles and predefined toolchains, creating an urgent need for flexible and decentralized coordination. Current structured communication protocols such as direct agent-to-agent messaging or MCP-style tool calls offer reliability, but they struggle to support the emergent and swarm-like intelligence required in large adaptive systems. Distributed agents must learn continuously, share context fluidly, and coordinate without depending solely on central planners.

This paper revisits gossip protocols as a complementary substrate for agentic communication. Gossip mechanisms, long valued in distributed systems for their decentralized and fault-tolerant properties, provide scalable and adaptive diffusion of knowledge and fill gaps that structured protocols alone cannot efficiently address. However, gossip also introduces challenges, including semantic relevance, temporal staleness, and limited guarantees on action consistency in rapidly changing environments.

We examine how gossip can support context-rich state propagation, resilient coordination under uncertainty, and emergent global awareness. We also outline open problems around semantic filtering, trust, and knowledge decay. Rather than proposing a complete framework, this paper presents a research agenda for integrating gossip into multi-agent communication stacks and argues that gossip is essential for future agentic ecosystems that must remain robust, adaptive, and self-organizing as their scale and autonomy increase.

\end{abstract}


\section{Introduction}

Agentic AI represents a new computational paradigm in which autonomous software agents perceive context, plan actions, invoke tools, and adapt their behaviour over time. These systems already underpin enterprise automation, scientific discovery, robotics, and large-scale orchestration tasks. As multi-agent deployments expand in scale and heterogeneity, the demands on communication infrastructures have intensified. Agents must continuously share partial knowledge, discover peers, negotiate roles, and align on the evolving environmental state — all while operating under uncertainty, incomplete information, and dynamic membership.

Current Multi-Agent Communication Protocols (MACPs), including the Model Context Protocol (MCP), the Agent Communication Protocol (ACP), and Google's A2A, provide essential foundations for structured interaction. These protocols support authenticated messaging, request--response exchanges, tool invocation, and inter-agent negotiation. However, they inherit a long-standing architectural assumption: that agent interactions must be explicit, directed, and serialised. This assumption hinders large-scale coordination, especially in environments characterised by failure, churn, or high information volatility.

A growing body of research argues that future agentic systems will behave less like isolated tools and more like adaptive collectives. Agents will dynamically refine their workflows, learn new capabilities, broadcast learned insights, and coordinate reflexively without centralised oversight. In these settings, knowledge does not merely flow along predetermined channels---it diffuses organically throughout the ecosystem. The communication substrate must therefore support emergent coordination, soft convergence, and distributed awareness.

Gossip protocols, widely used in distributed storage, peer-to-peer overlays, and service meshes, exhibit precisely these qualities. Their epidemic-style dissemination, stochastic peer selection, and redundancy-driven robustness allow systems to scale to thousands of nodes with minimal per-agent overhead. Despite their success in distributed systems, their role in agentic AI remains underexplored.

In this paper, we propose gossip as a \emph{substrate layer} for agentic communication. We examine how gossip integrates beneath MCP, A2A, and ACP to support decentralised discovery, load signalling, task-availability propagation, failure detection, and emergent consensus. Rather than replacing structured protocols, gossip augments them, providing the diffuse awareness necessary for large-scale autonomy.

To evaluate this claim, we present two complementary classes of benchmarks: (1) agentic benchmarks measuring discovery latency, load distribution, and semantic state propagation; and (2) domain benchmarks drawn from industrial automation and disaster response, illustrating performance under predictable and uncertain environments. The results illustrate that gossip is uniquely suited to the next generation of multi-agent systems: scalable, robust, adaptive, and capable of enabling collective intelligence beyond the limits of point-to-point communication.

\section{Background}

The evolution of multi-agent systems has been shaped by increasing needs for autonomy, scalability, and resilience. Early agent-based architectures emphasised symbolic communication and direct message passing \cite{wooldridge2009introduction}, while modern agentic AI integrates large language models (LLMs), tool invocation, and adaptive workflow generation. Across these paradigms, communication remains the primary enabler of collective intelligence.

\subsection{Structured Multi-Agent Communication Protocols}

Recent Multi-Agent Communication Protocols (MACPs), such as the Model Context Protocol (MCP) \cite{mcp2024}, Google's Agent-to-Agent (A2A) protocol \cite{a2a2024}, and the Agent Communication Protocol (ACP) \cite{acp2024}, provide structured foundations for authenticated message exchange, tool invocation, and semantic negotiation. These protocols ensure interoperability but inherit a directed, request--response communication model that constrains scalability in large or dynamic systems.

As agent deployments scale from tens to thousands, the assumption that all interactions must be explicit, synchronous, and addressed becomes increasingly limiting \cite{dafoe2020open}. The burden of explicit discovery, retries, and mediation hinders fluid coordination and runtime adaptability.

\subsection{Decentralization in Distributed Systems}

Distributed systems research approached these challenges from a different foundation. Gossip-based epidemic protocols \cite{demers1987epidemic, birman2007history} introduced stochastic, redundancy-driven communication that spreads information exponentially fast with negligible per-node cost. Gossip is now foundational in large-scale production systems, including Cassandra \cite{lakshman2010cassandra}, Dynamo \cite{decandia2007dynamo}, Consul \cite{consul2013}, Scuttlebutt \cite{brown2014scuttlebutt}, and Fireflies \cite{reifenberg2022fireflies}.

Key properties include:

\begin{itemize}
    \item \textbf{Epidemic dissemination} enabling logarithmic-time propagation \cite{kempe2003gossip}.
    \item \textbf{Constant per-node load} enabling horizontal scalability \cite{birman2011gossip}.
    \item \textbf{Fault tolerance} even under extreme churn \cite{renesse1998scalable}.
    \item \textbf{Emergent convergence} without central coordination \cite{almeida2003gossip}.
\end{itemize}

These align closely with the emerging requirements in large-scale agentic AI.

Growing demands for autonomy, scalability, and resilience have driven the evolution of multi-agent systems. Early agent architectures prioritized symbolic reasoning and explicit message passing \cite{wooldridge2009introduction}, while contemporary agentic AI increasingly leverages large language models (LLMs), dynamic tool invocation, and adaptive orchestration pipelines. Throughout this evolution, communication remains the foundational substrate enabling collective intelligence.

\subsection{Structured Multi-Agent Protocols and Their Limits}

Modern Multi-Agent Communication Protocols (MACPs) — such as the Model Context Protocol (MCP) \cite{mcp2024}, the Agent-to-Agent (A2A) protocol \cite{a2a2024}, and others — offer structured, authenticated, and interoperable channels for tool invocation, data access, and explicit agent coordination. These systems excel at ensuring deterministic messaging, capability invocation, and compliance with schema and security constraints. However, they share an inherent design assumption: communications resemble RPC (remote procedure call) or request-response interactions, with static peer directories, explicit discovery, and pre-configured workflows.

As agentic deployments scale into large or dynamic agent populations, these assumptions become increasingly constraining. Agents may join and leave, capabilities may evolve, and coordination needs may shift unpredictably. Structured MACPs place the burden of discovery, retries, and orchestration logic on developers — a burden that grows as system complexity increases. They struggle to accommodate fluid peer discovery, fault-tolerant broadcast of state, or emergent, context-driven coordination.

\subsection{Decentralized Gossip Protocols as an Alternative Substrate}

In distributed systems, gossip—or epidemic—protocols have long addressed the need for scalable, fault-tolerant dissemination of state across large populations of nodes. These protocols rely on randomized, peer-to-peer exchanges launched at regular intervals, ensuring that information eventually reaches every node with low per-node overhead and high resilience to churn or network partitions. Systems such as Cassandra, Dynamo, and Consul build on these mechanisms to maintain consistency, membership awareness, and failover handling across large clusters.

Recent work argues that gossip brings valuable—or even necessary—properties to future agentic AI systems. In the 2025 preprint “Revisiting Gossip Protocols: A Vision for Emergent Coordination in Agentic Multi-Agent Systems,” Habiba and Khan assert that gossip fills a critical gap in contemporary coordination architectures: enabling decentralized, low-overhead, context-rich communication that supports emergent, swarm-like behavior in large, dynamic agent populations \cite{habiba2025revisiting}. They observe that structured protocols, while reliable, are inadequate for reflexive, adaptive coordination, in which agents continuously learn, share context, and renegotiate roles outside pre-defined workflows.

Gossip, in this vision, functions as a complementary layer beneath or beside structured protocols—not as a replacement. It enables:

\begin{itemize}
    \item \textbf{Distributed awareness and peer discovery}, even with dynamic membership and evolving capabilities.
    \item \textbf{Low-overhead state diffusion and redundancy}, allowing soft state (e.g., resource load, task offers, context updates) to propagate with minimal per-agent cost.
    \item \textbf{Emergent coordination and consensus}, as agents gradually converge on shared context and make decentralized decisions informed by a broader, indirectly communicated knowledge base.
    \item \textbf{Fault tolerance and resilience}, maintaining global information flow despite node failure, partitioning, or communication loss.
\end{itemize}

\subsection{Towards Collective, Context-Rich Agentic Intelligence}

Agentic AI is increasingly moving toward environments in which agents are not constrained to fixed roles or workflows but must adapt, collaborate, and coordinate in open-ended, unpredictable contexts. In such settings, structured, deterministic communication may be insufficient or overly rigid. By contrast, gossip-based communication supports a more organic model of collaboration — one where agents continuously share, update, and reconcile their local knowledge and context without central orchestration.

The 2025 vision paper provides a roadmap for incorporating gossip as a first-class communication substrate in multi-agent ecosystems. It charts open challenges — semantic filtering, temporal staleness, trust and provenance, and conflict resolution — and proposes a research agenda to address them \cite{habiba2025revisiting}. This motivates our current work: to explore how gossip can be integrated into agentic orchestration patterns to enable distributed learning, reflexive adaptation, and resilient coordination at scale.

By combining structured MACPs for reliability and security with gossip for context-awareness and emergent coordination, we believe the next generation of agentic systems can achieve a robust balance: structured when needed, adaptive when beneficial, and scalable under uncertainty.

\subsection{Emergence and Collective Intelligence}

Future agentic ecosystems are expected to function as adaptive collectives in which agents share learned behaviour, broadcast discoveries, and autonomously negotiate roles \cite{yang2024agentverse}. Systems must therefore support:

\begin{itemize}
    \item distributed awareness,
    \item adaptive role negotiation,
    \item emergent task allocation,
    \item and collective situational convergence.
\end{itemize}

Swarm robotics research has demonstrated that gossip-like local interactions can yield robust global behaviour \cite{brambilla2013swarm, trianni2021review}. Yet gossip has not been systematically integrated beneath structured MACPs. This motivates a deeper investigation into gossip as a substrate for next-generation agentic AI.

\section{Problem Definition}

Despite advances in agent communication standards, modern agentic ecosystems exhibit structural limitations in scalability, resilience, and semantic alignment.

\subsection{Limitations of Directed Communication}

Directed protocols such as MCP, A2A, and ACP impose explicit addressing and point-to-point semantics \cite{mcp2024, a2a2024, acp2024}. These designs create several constraints:

\begin{itemize}
    \item \textbf{Limited runtime discovery:} agents cannot autonomously discover peers without a central registry \cite{yang2024agentverse}.
    \item \textbf{Centralization risks:} directories become single points of failure \cite{dafoe2020open}.
    \item \textbf{Fragility under churn:} dynamic joins or failures require explicit error-handling \cite{wooldridge2009introduction}.
    \item \textbf{Restricted awareness:} global state is not propagated unless explicitly queried.
\end{itemize}

These constraints hinder deployment in dynamic, open, or large-scale environments.

\subsection{Absence of Semantic-Relational Dissemination}

Most protocols focus on \emph{syntactic} structure, not \emph{semantics}. This introduces several gaps:

\begin{itemize}
    \item \textbf{No shared ontologies} for roles, goals, and capabilities \cite{gruber1995ontology}.
    \item \textbf{Shallow capability descriptions:} tools lack relational context, such as dependencies or prerequisites.
    \item \textbf{Missing provenance:} agents cannot reconstruct why actions were taken or how decisions relate to system context \cite{moreau2018provenance}.
    \item \textbf{Weak semantic cohesion:} agents struggle to situate their local knowledge within collective workflows.
\end{itemize}

This leads to brittle coordination across vendor ecosystems, platforms, and agent architectures.

\subsection{Need for Distributed, Emergent Coordination}

As agents become more autonomous, coordination must increasingly emerge from local interactions rather than explicit orchestration \cite{dafoe2020open, yang2024agentverse}. This requires:

\begin{itemize}
    \item decentralised task availability diffusion,
    \item failure detection without central monitors,
    \item runtime adaptation to environmental uncertainty,
    \item and organic formation of coalitions or swarms.
\end{itemize}

These capabilities align closely with epidemic-style communication but are not supported by existing MACPs.

\subsection{Substrate Gap}

Collectively, these limitations reveal a missing architectural layer: a decentralized, fault-tolerant communication substrate supporting:

\begin{itemize}
    \item asynchronous state diffusion,
    \item stochastic discovery,
    \item resilient awareness propagation,
    \item emergent coordination,
    \item and semantic-relational knowledge sharing.
\end{itemize}

Gossip protocols offer a promising foundation for this substrate by enabling scalable, low-overhead communication that complements rather than replaces structured protocols.

\section{Core Mechanics of Gossip in Agentic Systems}

Gossip protocols, long associated with large-scale distributed systems \cite{demers1987epidemic, birman2011gossip}, offer communication primitives that align remarkably well with the emerging demands of autonomous agent populations. While structured protocols such as MCP or A2A excel at typed, semantic message exchange, they rely on stable addressing and explicit intent. Gossip operates differently: it supports continual, redundant diffusion of state, enabling agents to approximate collective situational awareness without central orchestration. To establish this foundation, we first review the operational mechanics of gossip before integrating these components into the proposed GEACL substrate.

\subsection{Epidemic-Style Propagation}

Gossip dissemination follows an epidemic dynamic: each informed agent periodically selects a peer and exchanges state. The expected population-level growth is governed by logistic kinetics:

\[
    \frac{dI}{dt} = \beta I(t)\Big(1 - \frac{I(t)}{N}\Big),
\]

\begin{figure}[ht]
\centering
\includegraphics[width=0.9\linewidth]{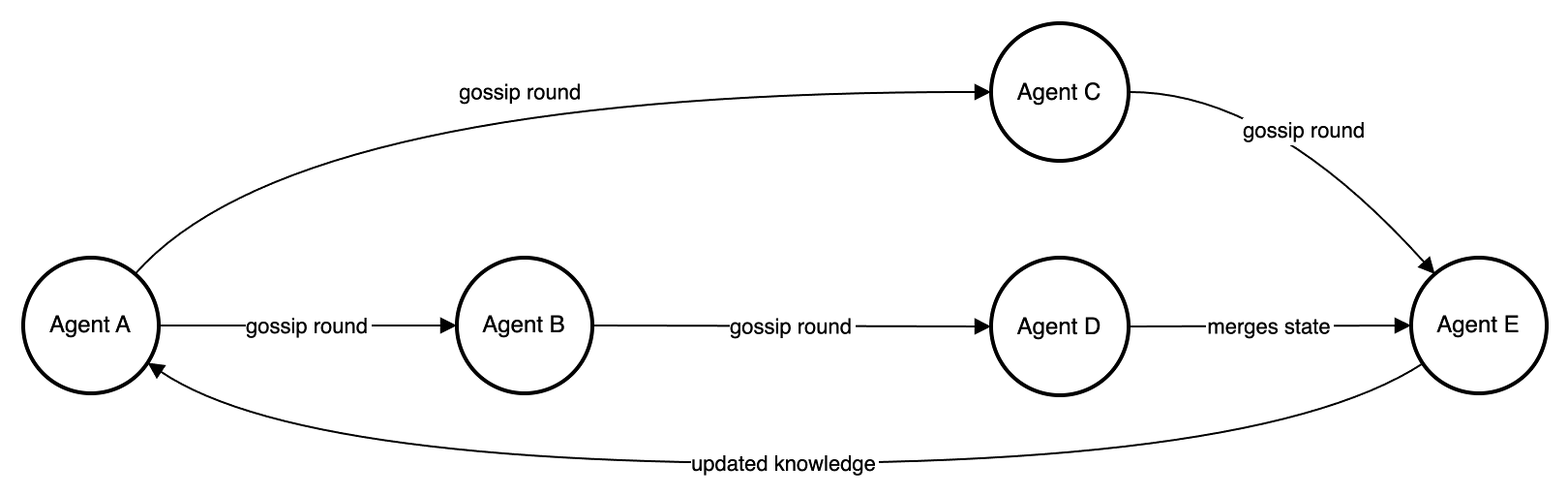}
\caption{Epidemic-style dissemination among agents. Each round expands the frontier of informed agents, producing eventual global convergence.}
\label{fig:gossip-epidemic}
\end{figure}

Where $I(t)$ is the number of agents aware of the update at time $t$ and $N$ is the population size. This produces the characteristic “S-shaped” convergence curve shown in Figure~\ref{fig:gossip-epidemic}, where propagation accelerates early and saturates as the system approaches complete dissemination. Significantly, the convergence time grows only logarithmically with $N$, enabling dissemination across thousands of agents with minimal coordination overhead \cite{kempe2003gossip}.

\subsection{Push, Pull, and Push--Pull Exchanges}

Building on epidemic mechanisms, gossip protocols differ primarily in the directionality of their exchange. The push model propagates updates aggressively from informed agents, whereas the pull model empowers uninformed agents to request missing knowledge. Push--pull combines these advantages, reducing worst-case latency and mitigating asymmetric state divergence. The transition to this subsection reflects an important point: while epidemic propagation drives dissemination speed, push--pull exchanges determine dissemination robustness—particularly when agents hold disjoint or partially overlapping knowledge fragments.

\subsection{Partial Views and Random Peer Sampling}

Propagation effectiveness depends critically on *which* peers agents contact. Rather than maintaining all-to-all connectivity, agents rely on peer-sampling protocols such as CYCLON or SCAMP \cite{almeida2003gossip}, maintaining a partial view $V_i(t)$ of limited size. This introduces locality, diversity, and churn-resilience into the system. The transition here is intentional: epidemic dynamics describe the *rate* of propagation, while peer sampling determines the *shape* of the propagation topology. Figure~\ref{fig:geacl-architecture} later illustrates how partial views form dynamic neighbourhoods that underpin decentralized coordination.

\subsection{Anti-Entropy and State Reconciliation}

Random exchanges inevitably introduce inconsistencies across snapshots among agents. Anti-entropy mechanisms resolve divergence through CRDT merges, vector clocks, or last-writer-wins rules. This transforms raw gossip into a semantic alignment mechanism: not only do agents share data, but they converge on a coherent representation of shared reality. Crucially, this introduces the concept of *semantic consistency*, which transitions naturally into rumour-mongering and prioritisation in the following subsection.

\subsection{Rumour-Mongering and Context-Aware Suppression}

Unbounded gossip propagation increases network load. Rumour-mongering reduces redundancy by suppressing updates once estimated coverage is sufficiently high. Unlike classical systems, agentic AI demands *context-aware suppression*: safety-critical updates (e.g., structural collapse in disaster response) must propagate aggressively, while routine updates (battery levels, minor task changes) can be throttled. This lays the conceptual groundwork for the GEACL framework, in which gossip is treated as a tunable substrate rather than a flat broadcast mechanism.

\subsection{Interpretation for Agentic AI}

Collectively, these mechanics form the operational basis for gossip in agentic ecosystems:

\begin{itemize}
    \item epidemic propagation provides scalable dissemination;
    \item push--pull exchanges enable mutual refinement of partial world models;
    \item partial views foster local collaboration at scale;
    \item anti-entropy ensures eventual semantic alignment;
    \item rumour suppression provides resource-aware prioritisation.
\end{itemize}

These components naturally complement structured multi-agent communication protocols. Figure~\ref{fig:geacl-architecture} integrates these elements into a unified substrate, forming the foundation for the Gossip-Enhanced Agentic Coordination Layer (GEACL) introduced in the next section.

\section{The Gossip-Enhanced Agentic Coordination Layer (GEACL)}

While structured protocols such as MCP, A2A, and ACP provide typed communication, authentication, and tool invocation, they do not inherently support decentralized discovery, ambient state diffusion, or emergent coordination. To address these gaps, we introduce the \textbf{Gossip-Enhanced Agentic Coordination Layer (GEACL)}, a substrate that integrates epidemic dissemination, semantic alignment, and stochastic peer engagement. GEACL provides the missing infrastructure for scalable, adaptive, self-organizing agentic ecosystems.

GEACL is designed around three core principles:

\begin{enumerate}
    \item \textbf{Diffuse globally, act locally:} agents maintain only local state but benefit from global information diffusion.
    \item \textbf{Coordinate without central orchestration:} system-level coherence emerges from stochastic interactions.
    \item \textbf{Augment, not replace, existing protocols:} gossip operates beneath semantic protocols, providing redundancy and fault tolerance.
\end{enumerate}

GEACL thus acts as a “communication substrate,” complementing explicit, structured interactions by enabling continual alignment of background states.

GEACL governs the decay of semantic divergence:
\[
D(t) = \sum_{i,j} d(S_i(t), S_j(t)),
\]
Where $d$ may include symbolic, relational, or embedding-based distance. Through push--pull gossip:
\[
D(t+1) \approx (1 - \eta) D(t),
\]
Where $\eta$ reflects fan-out, peer-sampling diversity, and merge semantics.

Convergence is guaranteed when $0 < \eta < 1$, and the convergence rate scales logarithmically with population size.

GEACL extends classical gossip through five augmented operations:

\begin{itemize}
    \item \textbf{Semantic Gossip:} agents exchange compressed state vectors, intentions, and capability metadata.
    \item \textbf{Contextual Prioritisation:} update propagation depends on event salience.
    \item \textbf{Adaptive Peer Sampling:} agents bias peer selection by proximity, expertise, or task overlap.
    \item \textbf{Merge-by-Meaning:} CRDT-like merges are enriched with semantic embedding models.
    \item \textbf{Background Synchronisation:} gossip runs concurrently with tool invocation and structured communication.
\end{itemize}
This provides a minimal-assumption substrate that enables emergent coordination without violating the semantics of upper-layer protocols.
\subsection{Architecture}

\begin{figure}[ht]
\centering
\includegraphics[width=0.65\linewidth]{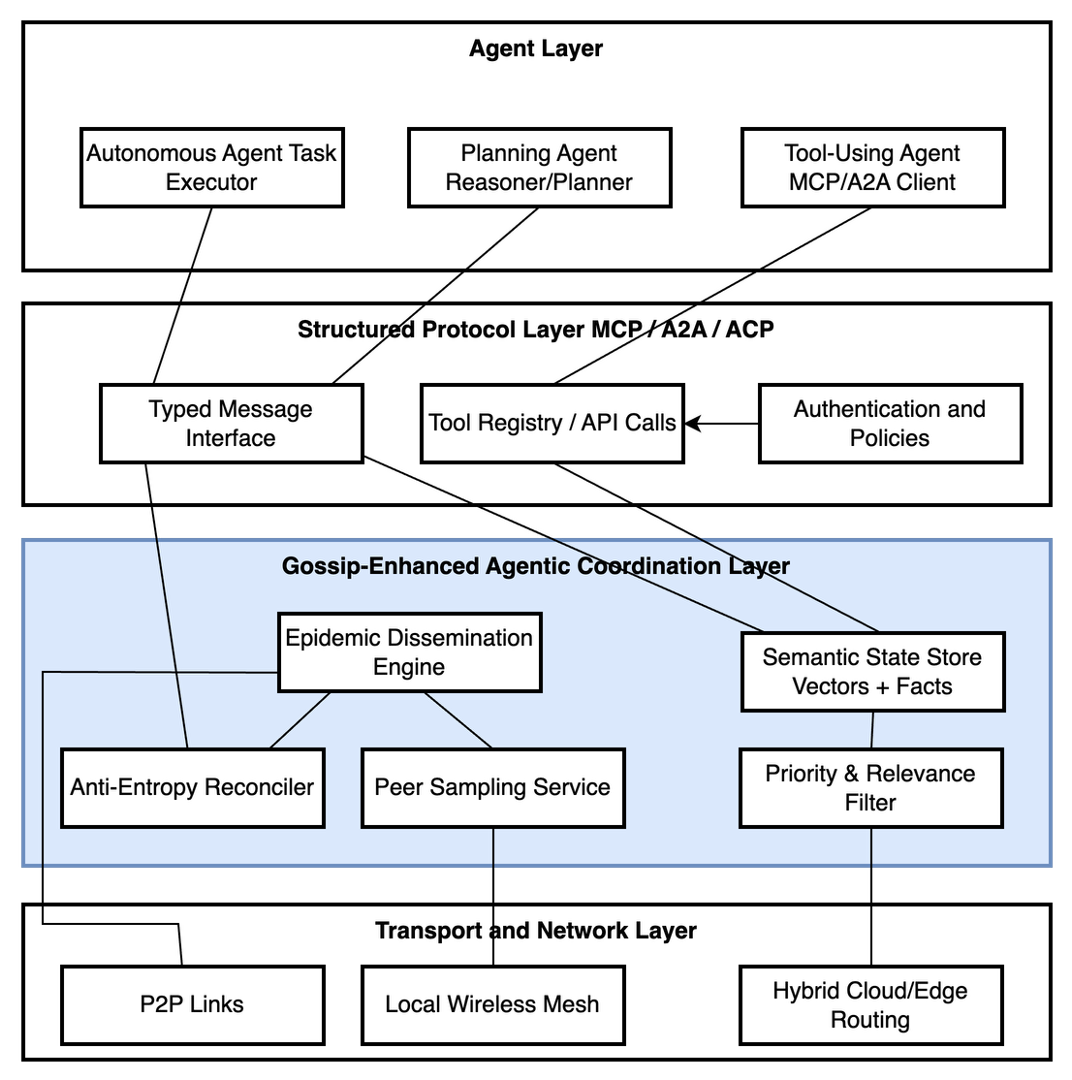}
\caption{GEACL: Architecture of the Gossip-Enhanced Agentic Coordination Layer.}
\label{fig:geacl-architecture}
\end{figure}

The GEACL architecture (Fig.~\ref{fig:geacl-architecture}) comprises four interacting layers. Each layer contributes a distinct functional capability, and together they form a decentralized substrate that supports scalable, adaptive multi-agent coordination. Unlike traditional pipeline-style diagrams, the architecture consists of loosely coupled, independently operating nodes that exchange information opportunistically through gossip while maintaining semantic precision through structured protocols. 

Here are detailed layers for the GEACL framework:

\subsubsection{Agent Layer}

The Agent Layer contains the autonomous entities responsible for reasoning, perception, planning, 
and task execution. Three representative agent types are shown for generality: (i) \textit{task executors} that directly operate in the environment, (ii) \textit{planning or reasoning agents} 
that manage task decomposition and decision-making, and (iii) \textit{tool-using agents} that 
interface with external APIs or structured communication protocols. These agents do not maintain global awareness directly; instead, they rely on GEACL to provide ambient knowledge about peer status, task availability, environment conditions, and system-level trends. In practice, the Agent Layer is the origin of intentions, observations, and queries that propagate downward to the communication layers.

\subsubsection{Structured Protocol Layer (MCP / A2A / ACP)}

This layer provides high-level semantic communication through well-defined message schemas such as 
JSON-RPC (MCP), directed messaging (A2A), or typed interaction patterns (ACP). Its role is to 
support \emph{explicit}, \emph{intent-driven}, and \emph{directional} communication. Typical 
operations include tool invocation, targeted information requests, and authenticated data exchange. 
However, this layer assumes the existence of stable addressing, pre-registered endpoints, and 
intentional message routing. It does not address decentralized discovery, background synchronization, 
or emergent consensus. Those elements are delegated to GEACL. Thus, the Structured Protocol Layer 
functions as the \emph{semantic messaging system}, but its effectiveness in dynamic, large-scale 
agent populations depend critically on the contextual information provided by the gossip substrate.

\subsubsection{Gossip-Enhanced Agentic Coordination Layer (GEACL)}

GEACL serves as the distributed substrate that maintains collective context across the agent 
ecosystem. It operates continuously and independently of explicit agent requests. The layer is 
composed of five cooperating components:

\begin{itemize}
    \item \textbf{Epidemic Dissemination Engine (G1).} Implements stochastic push--pull gossip, 
    allowing observations, intent traces, capability updates, and local beliefs to spread throughout 
    the system with logarithmic convergence time. It ensures that no single agent becomes a point of 
    failure for system-wide awareness.

    \item \textbf{Peer Sampling Service (G2).} Maintains a rotating partial view of neighbouring 
    agents using protocols such as CYCLON or SCAMP. This prevents the need for global membership 
    lists and ensures robust connectivity even under churn, failures, or ad-hoc deployments.

    \item \textbf{Semantic State Store (G3).} Each agent maintains a local store of symbolic facts, 
    embedding-based representations, task metadata, and environmental variables. The state exchanged 
    during gossip originates from this store and reflects an agent’s local understanding of the 
    system.

    \item \textbf{Anti-Entropy Reconciler (G4).} Applies conflict-resolution rules when agents hold 
    inconsistent knowledge. Merge functions may include CRDT joins, vector-clock ordering, or 
    semantic similarity alignment. This enables decentralized convergence of world models without 
    central arbitration.

    \item \textbf{Priority and Relevance Filter (G5).} Implements rumour-suppression and 
    event-weighting mechanisms. Safety-critical or time-sensitive updates are gossip-amplified, 
    whereas routine or low-value information is suppressed to prevent bandwidth saturation. This 
    context-aware filtering is essential for energy-limited or bandwidth-constrained deployments.
\end{itemize}

Whereas the Structured Protocol Layer handles high-precision directed communication, GEACL provides 
\emph{ambient synchronization} and \emph{collective awareness}. The interactions between G1--G5 
allow agents to maintain aligned world models, discover peers, and react adaptively to global 
conditions—all without centralized orchestration.

\subsubsection{Network Layer}

The Network Layer provides the physical transport for both structured messages and gossip 
exchanges. It may include peer-to-peer wireless links, ad-hoc mesh networks, or hybrid networks 
spanning cloud, fog, and edge resources. GEACL does not assume reliable connectivity; instead, it is 
designed to operate opportunistically across heterogeneous communication channels. This ensures that 
updates continue to propagate even during partial failures, network partitions, or degraded 
connectivity conditions.

\subsubsection{Cross-Layer Interaction}

The layers do not form a sequential pipeline but instead operate concurrently. For example, an agent 
may use MCP to query a tool while simultaneously receiving environmental updates through G1. 
Conversely, the state learned from gossip may influence which API calls the agent chooses to make. The 
interplay between layers creates a hybrid communication ecosystem: structured protocols ensure 
semantic precision and correctness, while GEACL ensures resilience, scalability, and decentralized 
coordination. Figures~\ref{fig:gossip-epidemic} and \ref{fig:geacl-architecture} illustrate how 
These interactions enable coherent system-level behaviour despite partial observability and 
stochastic communication patterns.

GEACL augments the structured protocol layer by continuously providing agents with updated knowledge about peers, environment, capabilities, and tasks.

\subsection{Toy Example: Decentralized Coordination in a Smart Factory}

To illustrate how the GEACL architecture functions in practice, we consider a
simplified smart-factory environment involving four heterogeneous agents:
a robotic assembly arm ($A_{\mathrm{arm}}$), a material-handling agent
($A_{\mathrm{mat}}$), a quality-inspection agent ($A_{\mathrm{q}}$), and a
planning agent ($A_{\mathrm{plan}}$). 
Each agent operates independently but depends on ambient information about
production flow, workstation health, and quality anomalies. Importantly, no
central controller orchestrates their behaviour.

\vspace{0.3em}
\noindent\textbf{Scenario:}
The quality-inspection agent detects an abnormal defect rate at Workstation~4.
This observation must be communicated to all other agents so they can adapt
their behaviour. Instead of relying on a centralized messaging service,
the update diffuses through the GEACL substrate.

\subsubsection*{Step 1: Local Observation and Semantic State Update}

The quality-inspection agent records the event in its local semantic store:
\[
S_{A_{\mathrm{q}}}(t)
\leftarrow 
S_{A_{\mathrm{q}}}(t) \cup
\{ \texttt{(defect\_spike, WS4, high\_severity)} \}.
\]
This update enters the Semantic State Store (G3), tagged as
\textit{high-priority}, ensuring G5 amplifies its dissemination.

\subsubsection*{Step 2: Gossip Dissemination Through GEACL}

At the next gossip interval $\tau$, 
the Epidemic Dissemination Engine (G1) selects a peer according
to the Peer Sampling Service (G2). 
For example, it may select $A_{\mathrm{mat}}$, and they engage
in a push--pull exchange:
\[
S_{A_{\mathrm{q}}} \leftrightarrow S_{A_{\mathrm{mat}}},
\]
followed by anti-entropy reconciliation through G4.

Because the event carries a high-severity flag, the Priority Filter (G5)
instructs G1 to propagate the update with minimal suppression.
Within a few rounds, the probability that any given agent remains uninformed
becomes small:
\[
\Pr[\text{uninformed at time } t]
\approx e^{-\beta t}.
\]

\subsubsection*{Step 3: Local Behavioural Adaptation}

Once $A_{\mathrm{arm}}$ receives the update, it reduces throughput:
\[
\texttt{speed}(A_{\mathrm{arm}}) \leftarrow 
0.8 \times \texttt{speed}(A_{\mathrm{arm}}).
\]

The material-handling agent reroutes arriving parts away from Workstation~4:
\[
\texttt{route}(A_{\mathrm{mat}}) \leftarrow \texttt{alternate\_path}.
\]

The planning agent updates its task graph, deprioritizing downstream tasks
that depend on WS4 until the anomaly resolves.

\subsubsection*{Step 4: Structured Protocol Interaction}

The planning agent uses the Structured Protocol Layer (e.g., MCP) to:
\begin{itemize}
    \item request diagnostic data from a maintenance microservice,  
    \item query historical logs from a database tool,  
    \item instruct a repair agent to evaluate the station.  
\end{itemize}

These tool calls are \emph{informed by} the gossip-propagated context.
GEACL did not replace MCP; rather, it provided the 
global situational awareness needed to decide which tools to invoke.

\subsubsection*{Step 5: Emergent Global Coordination}

Within seconds, the entire system exhibits coherent behaviour:
slowed assembly, rerouted materials, scheduled maintenance, and
task reprioritization. No message was broadcast globally, no registry
service was queried, and no agent directly coordinated all others.
Instead, coherence emerged from local interactions:
\[
\lim_{t \rightarrow \infty} 
\Delta(t) = 
\sum_{i,j} d(S_i(t),S_j(t))
\rightarrow 0,
\]
where $\Delta(t)$ is the semantic divergence across agents.

This example highlights how GEACL enables:
\begin{itemize}
    \item decentralized propagation of critical state updates,
    \item distributed semantic alignment through G4,
    \item adaptive behaviour triggered by ambient information,
    \item selective activation of structured protocol calls (MCP/A2A),
    \item and emergent coordination without a central controller.
\end{itemize}

Even in this simple scenario, GEACL transforms a set of independent
agents into a self-organizing collective capable of responding rapidly
to unexpected environmental changes.

\subsection{Why GEACL Enables Capabilities Missing Today}
Conventional multi-agent communication stacks---including MCP, A2A, and ACP---excel at 
structured message exchange but assume stable registries, explicit requests, and deterministic
tool invocation. These assumptions constrain autonomous behaviour in environments
characterised by partial observability, agent churn, heterogeneous capabilities, and
the need for emergent coordination. The proposed GEACL substrate addresses these
limitations by introducing a decentralised, continuously evolving communication fabric
that complements existing structured protocols. Its benefits arise from five
capabilities that are largely absent in current agentic systems:

\paragraph{(i) Runtime peer discovery without centralised registries.}
Current protocols require explicit registration (e.g., tool manifests in MCP or
capability descriptions in A2A). In contrast, GEACL’s gossip-driven peer sampling
enables agents to discover one another solely through decentralised exchanges.
This allows the agent population to expand, contract, or reorganise without
reconfiguration of any central service, thereby supporting open, dynamic ecosystems.

\paragraph{(ii) Distributed dissemination of task- and state-awareness.}
Rather than relying on explicit point-to-point messages or orchestrator-driven
notifications, GEACL enables ambient diffusion of relevant task cues,
environmental observations, and load information. This allows agents to become
aware of system conditions (e.g., overloads, failures, available tasks) 
in the absence of direct requests, supporting reflexive coordination
analogous to biological collectives.

\paragraph{(iii) Gradual semantic convergence through anti-entropy reconciliation.}
Modern agentic systems lack mechanisms for agents to align their internal
representations of the environment beyond deliberate API calls. GEACL provides
a decentralised reconciliation process whereby agents merge partial world
models over time. This continuous alignment of semantic state supports
collective situational awareness without imposing strict consistency
requirements or synchronisation barriers.

\paragraph{(iv) Support for emergent self-organisation in heterogeneous populations.}
Because gossip propagation is stochastic, redundant, and topology-agnostic,
GEACL enables the formation of locally coherent behavioural patterns (e.g.,
task partitioning, spatial coverage, workload balancing) without global
planning. Such emergent organisation is a prerequisite for large-scale
agentic systems that exhibit swarm-like adaptability.

\paragraph{(v) Robust knowledge propagation under failure and network degradation.}
Structured protocols assume reliable connectivity or the availability of
designated authority nodes. GEACL, by contrast, maintains dissemination
despite failures, partitions, or partial outages. Knowledge continues to
flow through surviving communication paths, allowing agents to recover
system state and continue coordinated behaviour even under adversarial or
resource-constrained conditions.

\medskip
Taken together, these capabilities position GEACL as a missing substrate layer
that bridges the gap between deterministic, tool-centric communication and
the decentralised knowledge diffusion required for autonomous multi-agent
intelligence. Rather than replacing existing protocols, GEACL augments them
by providing the background informational fabric necessary for scalable,
adaptive, and resilient coordination.

\section{Trust, Noise, and Secure Gossip Extensions}

While gossip provides a scalable and fault-tolerant substrate for distributed coordination, its decentralised design makes it inherently vulnerable to noise, staleness, and adversarial behaviour. In contrast to structured protocols such as MCP, A2A, or ACP---which mediate interactions through authenticated, intent-driven channels, epidemic dissemination assumes that each node is an honest, cooperative participant. This assumption breaks down quickly in open multi-agent ecosystems, where agents may be misaligned, faulty, or compromised. 

In this section, we analyse the core challenges associated with trust and information quality in gossip-based communication, and explore emerging extensions that allow gossip systems to remain robust while preserving their decentralised nature.

\subsection{Propagation of False or Malicious Information}

A central risk of gossip is its amplifying effect: once a false statement enters the network, the epidemic mechanism rapidly spreads it to a significant fraction of agents. Unlike centralised systems, where a single authority can validate or revoke updates, gossip lacks a canonical source of truth. Consequently, even a single malicious agent can influence global behaviour if updates are not validated.

Examples include:
\begin{itemize}
    \item adversarial agents repeatedly gossiping inflated workload to avoid task assignments,
    \item drones broadcasting false hazard locations in a disaster zone,
    \item compromised software agents injecting bogus capability descriptors, misleading task delegation,
    \item or multi-agent financial trading bots spreading outdated or contradictory market states.
\end{itemize}

Such disruptions emerge not from sophisticated attacks but from inherent protocol properties: redundancy ensures propagation; lack of ordering allows overwriting; and local trust assumptions give adversaries leverage. 

\subsection{Message Amplification and Noise Accumulation}

Even honest agents can unintentionally degrade information quality. Due to repeated retransmission, stale or low-value updates may dominate the communication budget unless explicit \emph{attenuation} or \emph{suppression} mechanisms are employed. Noise accumulation is particularly harmful in:
\begin{itemize}
    \item dense networks with high fan-out,
    \item environments with rapidly changing state (e.g., mobility, volatile sensors),
    \item heterogeneous agent populations with inconsistent measurement fidelity.
\end{itemize}

Without constraints, gossip may overwhelm bandwidth, drain battery-powered agents, or produce inconsistent converged states.

\subsection{Staleness and Conflict Resolution}

Traditional gossip protocols assume monotonic state evolution (e.g., membership lists, binary health status). Agentic AI systems, however, exchange high-dimensional state and intent, which evolve non-monotonically. Without explicit mechanisms to manage:
\begin{itemize}
    \item update ordering,
    \item expiration,
    \item and conflict reconciliation,
\end{itemize}
Agents may converge on inconsistent or outdated knowledge.

Version vectors, hybrid logical clocks, or CRDT-style merge functions can mitigate these issues, though applying them to semantic agent state remains an open research direction.

\subsection{Secure Gossip Through Cryptographic Authenticity}

A foundational extension for trustworthy gossip is adding \emph{authenticity guarantees}:

\begin{itemize}
    \item \textbf{Digital signatures} ensure that updates originate from legitimate agents.
    \item \textbf{Certificates or capability tokens} embed permissions directly into gossiped messages.
    \item \textbf{Encrypted payloads} restrict access to sensitive portions of the state.
\end{itemize}

These measures provide a cryptographic backbone similar to that used in secure pub/sub or zero-trust service meshes, but in a fully decentralised setting. When combined with identity attestation, they significantly reduce impersonation and tampering.

\subsection{Reputation and Score Propagation}

Beyond authenticity, agents must assess \emph{credibility}. Reputation gossip---where nodes distribute compact trust scores alongside state---allows agents to discount updates from unreliable peers. Such approaches mirror biological systems, where organisms weigh signals based on source reliability.

Reputation diffusion enables:
\begin{itemize}
    \item down-weighting suspected nodes,
    \item opportunistic trust-building in dynamic networks,
    \item and collaborative filtering of anomalous behaviour.
\end{itemize}

However, reputation itself becomes subject to gossip's probabilistic dynamics; stabilising these meta-signals requires careful design to avoid cascading false distrust.

\subsection{Redundant Validation Through Multi-Source Corroboration}

A robust technique is \emph{majority corroboration}: agents act on new information only after receiving it from multiple independent peers. This mitigates one-shot poisoning and aligns with Bayesian consensus models, in which convergent evidence increases confidence.

This mechanism is particularly valuable in agentic AI for:
\begin{itemize}
    \item verifying environmental hazards in robotics,
    \item cross-checking intent signals during distributed planning,
    \item confirming the availability or failures of peers,
    \item and filtering hallucinated or noisy perceptual data.
\end{itemize}

Corroboration increases latency slightly but significantly enhances reliability in environments with untrusted agents. In addition, using signatures among agents, as shown in Fig.~\ref{fig:trusted-gossip}, can also improve the trust among them.

\begin{figure}[!htb]
\centering
\includegraphics[width=0.65\linewidth]{figs/02.png}
\caption{Trusted gossip propagation. Agents apply signatures for authenticity and rely on multi-source corroboration before committing to actions.}
\label{fig:trusted-gossip}
\end{figure}

\subsection{Implications for Agentic AI}

Secure gossip extends epidemic communication from simple replication to a resilient substrate capable of supporting high-level behavioural coordination. It enables:
\begin{itemize}
    \item decentralised trust formation,
    \item noise-tolerant consensus,
    \item adaptive filtering of unreliable information,
    \item and robust coordination even in adversarial or unstable settings.
\end{itemize}

These capabilities are essential for large-scale, heterogeneous agent societies where no single communication protocol, directory, or orchestrator can guarantee correctness or continuity. Gossip does not replace structured protocols such as MCP or A2A; instead, it complements them by providing a background layer of situational alignment and system-wide resilience.

\section{Agentic Use Cases for Gossip}

Gossip protocols support a range of coordination behaviours that arise naturally in large, dynamic populations of agents. Unlike explicit messaging, which requires directed intent and addressing, gossip enables spontaneous, emergent exchange of state, intent, and capability information. This section highlights several key use cases in which epidemic-style communication offers distinct advantages for modern agentic AI systems.

\subsection{Task Availability Dissemination}

In autonomous multi-agent environments, tasks may appear unpredictably and may be relevant only to a subset of agents. Directed request--response approaches require agents to poll central registries or rely on pre-specified orchestrators, creating latency and bottlenecks. 

Gossip enables decentralised task diffusion: when an agent encounters a new task, it gossips a minimal descriptor to a randomly selected subset of peers. Over several rounds, the task information permeates the population, allowing any relevant agent to claim it. The process is analogous to biological recruitment behaviours, such as waggle dances in bees or pheromone trails in ants, in which information propagates implicitly through local interactions.

This yields flexible and scalable task allocation without a global scheduler, especially valuable in systems with high churn or intermittent connectivity.

\subsection{Peer Load Signalling and Adaptive Balancing}

In large-scale agent collectives, achieving balanced utilisation requires each agent to maintain awareness of the workload distribution. Traditional load-balancing techniques rely on centralized controllers or explicit monitoring queries, which introduce fragility.

With gossip, each agent periodically shares lightweight metadata about its workload (e.g., queue length, energy level, current commitments). The resulting softly consistent global view enables agents to:

\begin{itemize}
    \item route new tasks toward underloaded peers,
    \item redistribute responsibilities,
    \item and avoid hotspots or cascading overload.
\end{itemize}

The emergent behaviour resembles homeostasis in biological systems, where global equilibrium arises from local exchanges rather than explicit coordination.

\subsection{Capability Advertisement and Evolution}

As agents acquire new tools, skills, or models, other agents must discover these capabilities to form effective coalitions. Current MACPs support capability invocation but require explicit queries or static registration.

Gossip enables \emph{capability diffusion}: agents periodically broadcast compact descriptors (e.g., ``supports summarisation,'' ``can control camera drone,'' ``has thermal imaging''). Over time, the population converges on a shared capability map without a registry. 

This is particularly powerful in:

\begin{itemize}
    \item heterogeneous swarms with diverse sensors,
    \item mixed fleets of physical and digital agents,
    \item open ecosystems where agents join and leave continuously.
\end{itemize}

Gossip thereby forms a substrate for emergent specialization and adaptive role assignment.

\subsection{Failure Detection and Fault Recovery}

Failure detection is a canonical application of gossip in distributed systems \cite{renesse1998scalable, lakshman2010cassandra}. Agentic AI inherits the same need: agents may go offline, encounter errors, or become isolated due to environmental constraints.

With gossip-based heartbeat dissemination, failure information spreads naturally through the population, enabling:

\begin{itemize}
    \item rapid local awareness of peer drop-out,
    \item dynamic reallocation of tasks held by failed peers,
    \item provisional suspicion and confirmation stages,
    \item decentralised reconfiguration of sub-teams.
\end{itemize}

This avoids reliance on central monitors and preserves operational continuity under severe uncertainty, especially in physical domains such as robotics or disaster response.

\subsection{Intent and State Propagation}

A critical challenge for agentic AI is preventing conflicting actions when multiple agents respond independently to similar stimuli. Gossip provides a lightweight means of diffusing \emph{intent information}: agents exchange not only their states but also their planned actions.

Example: if a robot intends to inspect a particular region, it gossips its plan. Peers may then decide to avoid redundant effort or coordinate complementary activities. This mirrors coordination mechanisms observed in collective animal behaviour, in which individuals broadcast movement intentions or risk assessments.

Intent gossip enables agents to form loose consensus patterns without strict synchronisation, allowing the emergence of cohesive group behaviour in open environments.

\begin{figure}[!htb]
\centering

\includegraphics[width=0.75\linewidth]{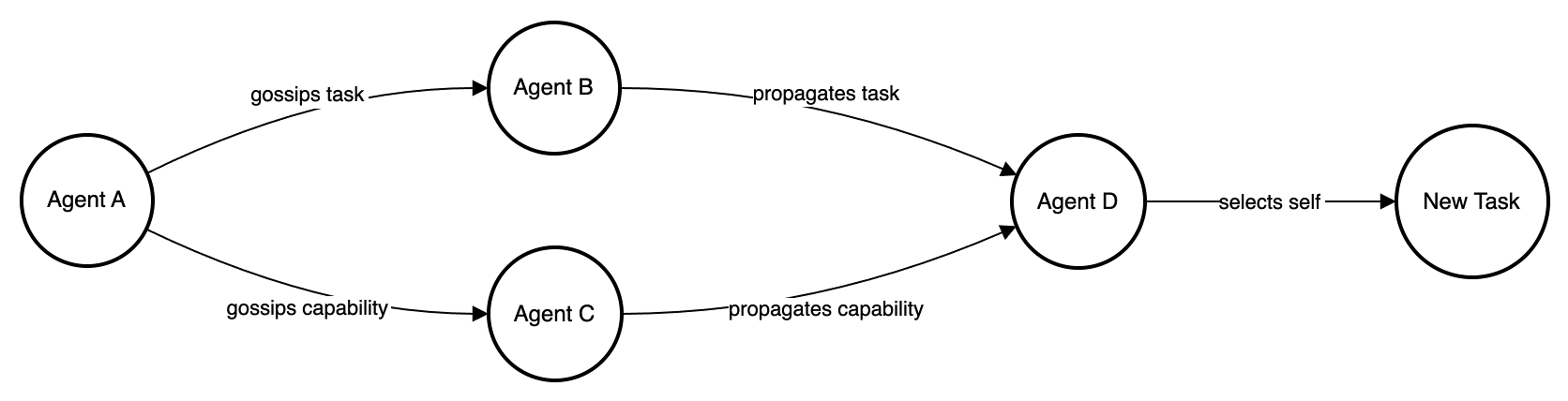}

\caption{Gossip-based dissemination of tasks and capabilities. Agents independently propagate and act on emerging opportunities, enabling decentralized task allocation.}
\end{figure}

\subsection{Interpretation in Agentic AI}

The above use cases illustrate how gossip provides:

\begin{itemize}
    \item \textbf{soft coordination}, avoiding the brittleness of explicit orchestration,
    \item \textbf{scalable awareness}, essential for large populations,
    \item \textbf{implicit consensus}, emerging from repeated local interactions,
    \item \textbf{adaptive behaviour}, resilient to uncertainty and partial failure.
\end{itemize}

These properties uniquely complement MACPs such as MCP and A2A, which excel at targeted, semantic, tool-oriented messaging but lack a background substrate for collective situational alignment.

Together, directional protocols and gossip form a hybrid architecture: the former governing explicit intent, the latter enabling emergent coordination and shared awareness.

\section{Design Patterns and Architectural Variants}

Gossip protocols exhibit multiple operational variants—push, pull, hybrid, anti-entropy, CRDT-based, and semantically filtered dissemination. In traditional distributed systems, these variants optimize convergence, bandwidth efficiency, and fault tolerance \cite{birman2007the, decandia2007dynamo}. 
In agentic AI, however, the key question is *how these variants map to established orchestration patterns* such as Swarm, Federated, Blackboard, Ecosystem, and Dependency-Driven workflows. 
Table~\ref{tab:designvariants} synthesizes this mapping, providing a pattern-centric rather than mechanism-centric view.

\begin{table*}[ht]
\centering
\caption{Mapping of Gossip Variants to Agentic AI Orchestration Patterns}
\label{tab:designvariants}
\renewcommand{\arraystretch}{1.25}
\begin{tabularx}{0.95\textwidth}{@{}l X X@{}}
\toprule
\textbf{Gossip Variant} & \textbf{Mechanism Summary} & \textbf{Agentic Pattern(s) Enabled} \\
\midrule

Push Gossip &
Updates proactively disseminated to neighbours; fast fan-out \cite{demers1987epidemic}. &
Swarm Coordination; Federated Discovery; Ecosystem Awareness. \\

Pull Gossip &
Nodes request missing or newer information, increasing convergence guarantees. &
Dependency-Driven Readiness Checking; Selective Role Delegation. \\

Hybrid Push–Pull &
Combines rapid propagation with robust convergence; baseline in many distributed systems \cite{jung2008efficient}. &
Ecosystem Patterns with heterogeneous agents; cross-domain capability discovery. \\

Anti-Entropy (State Repair) &
Nodes periodically compare digests (e.g., Merkle trees) to reconcile divergence \cite{decandia2007dynamo}. &
Blackboard Knowledge Bases; Shared World Models; Long-running Federated Clusters. \\

CRDT-Backed Gossip &
Conflict-free replicated structures ensure deterministic merges under concurrency \cite{shapiro2011conflict}. &
Distributed Blackboards; Symbolic or Relational Knowledge Convergence. \\

TTL / Semantic Filtering &
Messages decay based on age or relevance, making them useful for high-dimensional agent states \cite{valente2023distributed}. &
Swarm and Supply-Chain Patterns; Context-Bound Workflow-as-Knowledge propagation. \\

\bottomrule
\end{tabularx}
\end{table*}

\begin{figure}[!htb]
\centering
\includegraphics[width=0.75\linewidth]{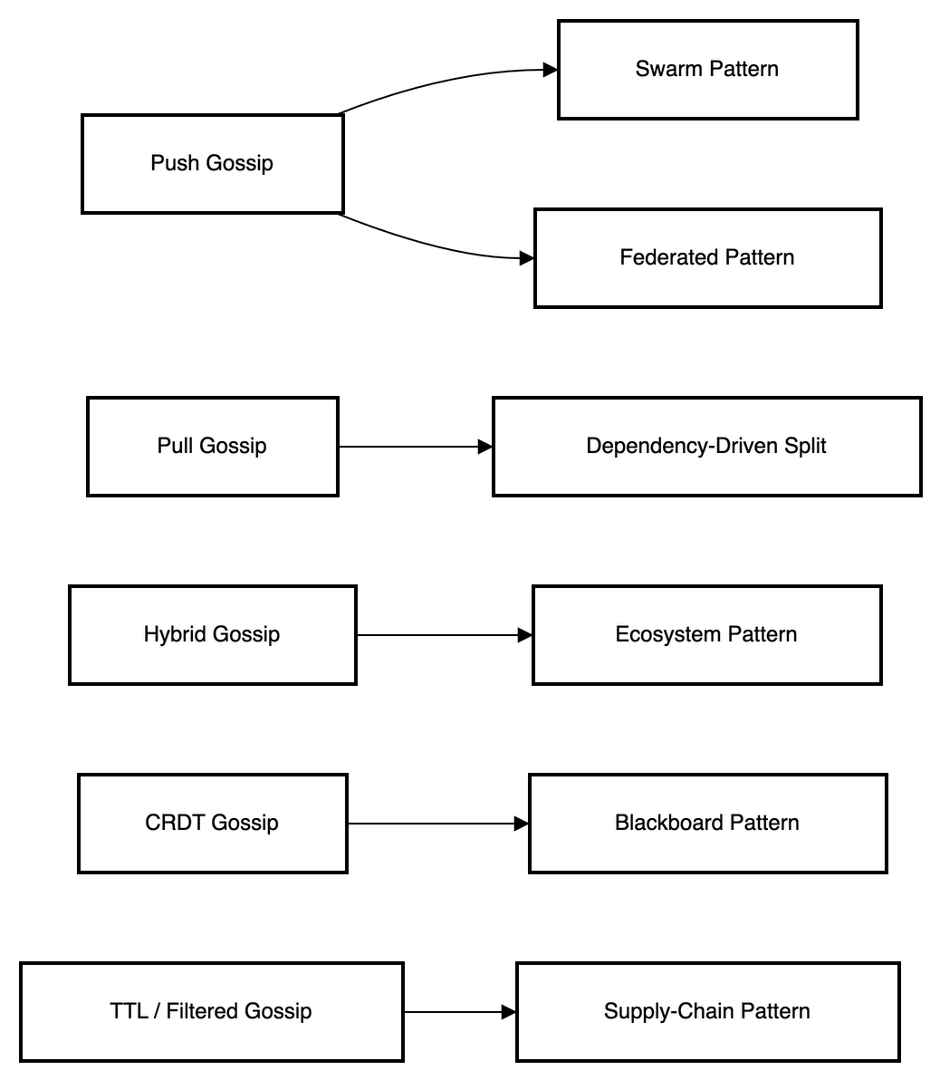}
\caption{High-level mapping of gossip communication variants to agentic orchestration patterns.}
\end{figure}

A pattern-centric analysis shows that gossip is not a monolithic protocol but a \emph{family of dissemination strategies} that align differently with agentic workflows:

\begin{itemize}
    \item \textbf{Push-based variants} benefit scenarios requiring rapid state diffusion—ideal for swarms, federations, and large agent ecosystems.
    \item \textbf{Pull and hybrid gossip} suit structured agent groups where readiness, role allocation, or dependency satisfaction must converge with bounded inconsistency.
    \item \textbf{CRDT and anti-entropy mechanisms} enable persistent distributed blackboards—critical for collective memory in heterogeneous agent clusters.
    \item \textbf{TTL and semantic filtering} address one of the key gaps in agentic systems: bandwidth-efficient dissemination of high-dimensional internal states.
\end{itemize}

This reframing avoids redundancy and highlights the architectural significance of gossip as a \emph{design palette} rather than a single mechanism. It clarifies where gossip adds robustness or autonomy to agentic AI patterns, and where its mathematical properties (e.g., eventual consistency, monotonic merges) can compensate for limitations in today’s RPC-style protocols (MCP, A2A, ACP).

\section{When Gossip Works---and When It Does Not}

Gossip protocols are robust but not universally applicable. Their strengths and limitations are well-studied in distributed systems \cite{birman2007the}, peer-to-peer overlays \cite{jesus2015survey}, and swarm robotic coordination \cite{brambilla2013swarm}. Agentic AI inherits these properties, but with richer semantic demands.

\subsection{Where Gossip Works Well}

\paragraph{Dynamic Discovery and Membership.}
In large, open ecosystems, gossip-based membership management avoids single points of failure \cite{das2002gossip}. This aligns with Ecosystem and Federated patterns.

\paragraph{Emergent Coordination and Load Sharing.}
Swarm behaviors can be replicated through epidemic information exchange, enabling task allocation, exploration, and role emergence \cite{campo2010task}.

\paragraph{Resilience and Failure Detection.}
SWIM demonstrates $O(1)$ failure detection and $O(\log n)$ dissemination \cite{das2002gossip}. Agentic systems benefit from decentralised fault tolerance without a central heartbeat monitor.

\paragraph{Collective Knowledge Maintenance.}
Distributed blackboards converge reliably when backed by gossip and CRDTs \cite{shapiro2011conflict}.

\subsection{Where Gossip Falls Short}

\paragraph{Strict Ordering.}
Applications requiring deterministic sequencing (e.g., financial robotics, industrial automation) need consensus \cite{ongaro2014raft}, not epidemic propagation.

\paragraph{Immediate Consistency.}
Real-time multi-agent control in safety-critical scenarios cannot afford the inherent latency and inconsistency of gossip \cite{valente2023distributed}.

\paragraph{Confidential or Sensitive Information.}
Broadcast-style dissemination requires nontrivial cryptographic protection to prevent leakage or Sybil attacks \cite{liu2020survey}.

\paragraph{Auditable and Deterministic Workflows.}
Hierarchical and legally governed agent workflows require predictable, traceable communication patterns \cite{lakemeyer2018artificial}.

\subsection{Hybrid Architectures}

Recent multi-agent research advocates integrating gossip as a background awareness substrate beneath structured RPC-like protocols \cite{mayer2024semantic}. MCP/A2A handle intent, roles, and tools; gossip maintains weakly consistent situational alignment.

\begin{figure}[ht]
\centering
\includegraphics[width=0.75\linewidth]{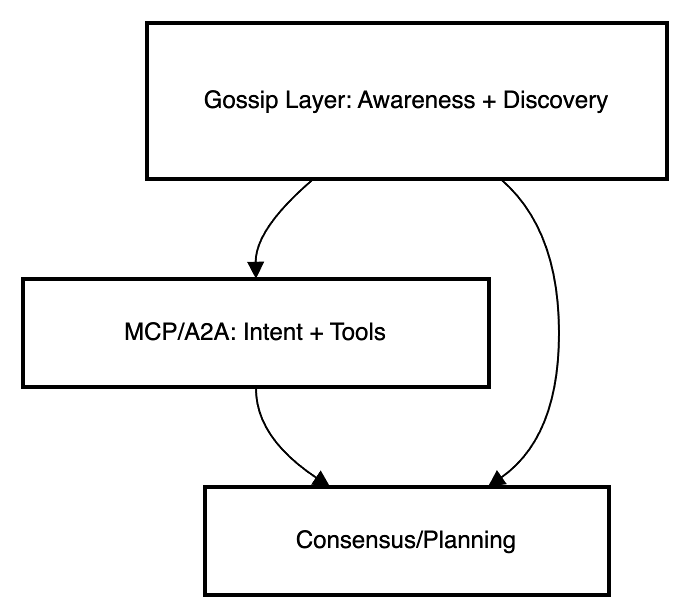}
\caption{Hybrid communication stack in multi-agent systems.}
\end{figure}

\section{Benchmarking and Evaluation Frameworks}

Evaluating gossip-enhanced agentic communication requires metrics and methodologies that differ significantly from those used for classical distributed systems or contemporary LLM-based agent workflows. 
Unlike deterministic RPC-style coordination, gossip-based interaction introduces stochasticity, partial visibility, and asynchronous convergence. 
Thus, a rigorous evaluation framework must quantify not only performance and scalability, but also emergent coordination quality, robustness under uncertainty, semantic propagation fidelity, and cross-agent relational consistency.  
This section formalizes such a framework and synthesizes benchmarking practices from distributed systems, swarm robotics, MARL, and multi-agent knowledge systems.

\subsection{Convergence and Propagation Metrics}

Gossip performance is primarily governed by convergence properties—how quickly and reliably information diffuses through an agent population.  
Foundational epidemic theory \cite{demers1987epidemic, shah2009gossip} shows that dissemination time follows a logarithmic relation with population size.  
For agentic systems, we define:

\begin{itemize}
    \item \textbf{Propagation Latency (PL):} Expected number of gossip rounds until 95\% of agents receive a fact.
    \item \textbf{Full Convergence Time (FCT):} Rounds until all non-failed agents converge; analogous to the "infection time" in SI/SIR models \cite{pastor2001epidemic}.
    \item \textbf{Divergence Window (DW):} Period during which agents hold inconsistent views; crucial for safety-critical coordination.
    \item \textbf{Propagation Coverage (PC):} Fraction of agents reached under partitioned or lossy connectivity.
\end{itemize}

These metrics provide quantitative measures for the suitability of gossip as a communication substrate in multi-agent workflows requiring implicit shared context.

\subsection{Robustness and Fault Tolerance}

Robustness is a key advantage of gossip systems, particularly in dynamic agent ecosystems where nodes fail, join, or move unpredictably.  
We adopt well-established distributed-systems benchmarks \cite{decandia2007dynamo, gupta2001capacity}:

\begin{itemize}
    \item \textbf{Failure Propagation Delay (FPD):} Time until agents become aware of a failure; SWIM achieves $O(1)$ detection time \cite{das2002swim}.
    \item \textbf{Availability Under Churn (AUC):} Fraction of correct updates delivered under high join/leave rates.
    \item \textbf{Network Partition Resilience (NPR):} Fraction of messages recovered after re-connection in delay-tolerant mobility scenarios.
    \item \textbf{Redundancy Overhead (RO):} Ratio of total messages sent versus minimum required for convergence.
\end{itemize}

Such metrics directly correspond to real agentic environments—particularly open ecosystems (IoT, drone swarms), where stochastic failures and topology shifts are the norm.

\subsection{Semantic and Relational Consistency Metrics}

One of the central gaps in current agent communication is the lack of semantic coherence across agents.  
We therefore introduce metrics assessing relational knowledge propagation and semantic fidelity:

\begin{itemize}
    \item \textbf{Semantic Convergence Score (SCS):} KL divergence or cosine similarity between relational embeddings held by each agent, inspired by work on distributed knowledge graphs \cite{jiang2021distributed}.
    \item \textbf{Capability Graph Alignment (CGA):} Consistency of agent capability maps gossiped across peers (e.g., which agent can do what task).
    \item \textbf{Relational Drift Index (RDI):} Degree of structural inconsistency in shared task-taxonomies or dependency graphs.
    \item \textbf{Intent Consistency Rate (ICR):} Percentage of agents whose internal plan representations align after gossip-based intent diffusion \cite{mahajan2023intent}.
\end{itemize}

These metrics allow quantitative evaluation of a gossip-enhanced semantic-relational substrate—something absent from RPC-based systems like MCP or A2A.

\subsection{XI-D. Emergent Coordination and Task Allocation Benchmarks}

Inspired by swarm robotics \cite{brambilla2013swarm} and MARL evaluation suites \cite{lowe2017multi}, we consider:

\begin{itemize}
    \item \textbf{Distributed Task Completion Rate (DTCR):} Fraction of tasks completed without centralized orchestration.
    \item \textbf{Redundant Action Minimization (RAM):} How well gossip reduces duplicate actions through emergent awareness.
    \item \textbf{Adaptive Load Rebalancing Effectiveness (ALRE):} Reduction in idle time or bottlenecks due to load gossip.
    \item \textbf{Coverage Efficiency (CE):} Spatial or logical area collectively explored without explicit planning (key in disaster response \cite{vahdat2000epidemic}).
\end{itemize}

This forms the basis for evaluating the effectiveness of gossip-enabled agentic orchestration patterns (Swarm, Ecosystem, Federated).

\subsection{Communication Cost and Efficiency}

Communication efficiency remains critical for practical deployment, especially in bandwidth-limited or energy-constrained settings.  
Benchmarks include:

\begin{itemize}
    \item \textbf{Messages Per Agent Per Round (MPAR)} and network usage scaling.
    \item \textbf{Bandwidth Efficiency Ratio (BER)} comparing gossip vs. direct RPC.
    \item \textbf{Energy Consumption per Dissemination Cycle (ECDC)} in mobile agents (drones, ground robots).
    \item \textbf{Selective Dissemination Efficiency (SDE)} for semantically filtered gossip \cite{valente2023distributed}.
\end{itemize}

These metrics ensure gossip protocols remain feasible in real-time physical systems, not just in cloud-hosted agents.

\subsection{Benchmarking Methodology}

We propose a multi-layer evaluation pipeline for agentic systems enhanced with gossip as shown in Fig.~\ref{fig:benchmark_method}:

\begin{figure}[!htb]
    \centering
    \includegraphics[width=0.65\linewidth]{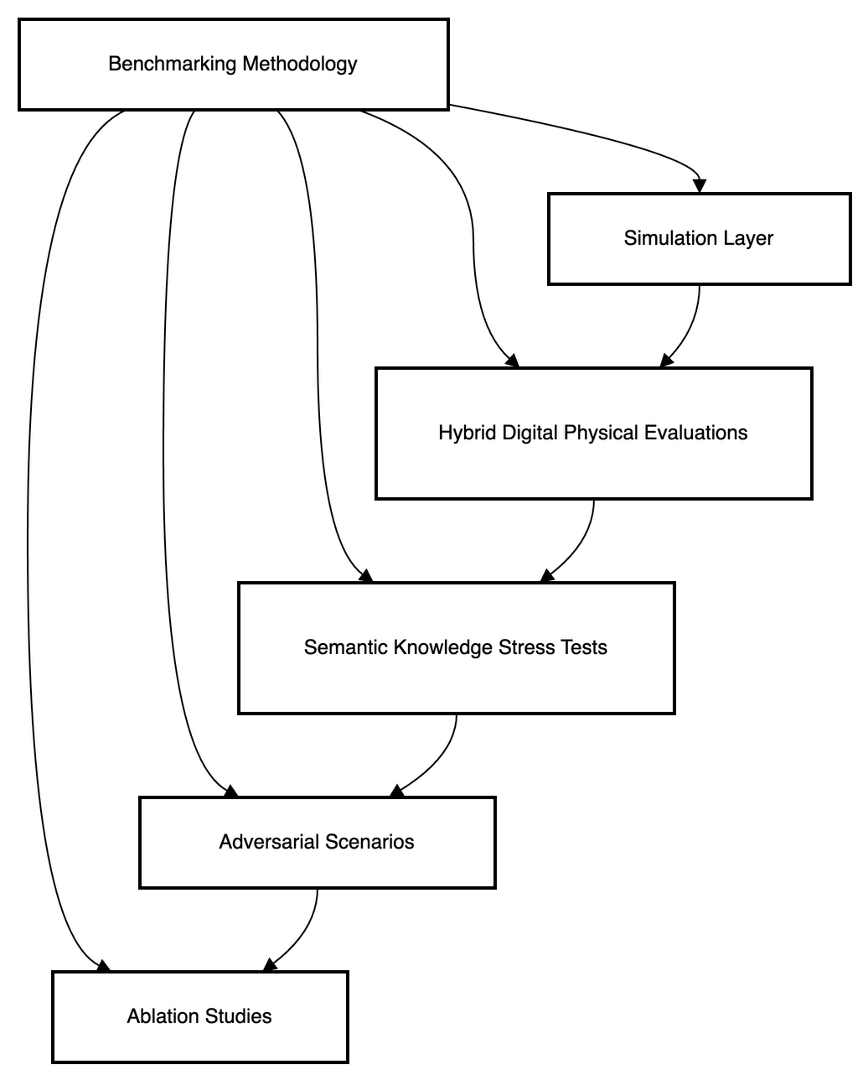}
    \caption{Proposed Benchmarking Methodology}
    \label{fig:benchmark_method}
\end{figure}

\begin{enumerate}
    \item \textbf{Simulation Layer}: Large-scale controlled experiments using agent simulators (e.g., Ray, MARLlib, Webots).
    \item \textbf{Hybrid Digital–Physical Evaluations}: Partial deployment on physical robots or IoT devices to test partition resilience.
    \item \textbf{Semantic Knowledge Stress Tests}: Injecting high-dimensional data (intent, embeddings, capabilities) to measure semantic convergence.
    \item \textbf{Adversarial Scenarios}: Testing robustness against false information injection or Sybil-style identity attacks \cite{douceur2002sybil}.
    \item \textbf{Ablation Studies}: Comparing gossip-only, RPC-only (A2A/MCP), and hybrid architectures.
\end{enumerate}

This holistic methodology evaluates gossip not merely as a dissemination mechanism but as a foundational coordination substrate for future agentic systems.

\subsection{Toward Standard Benchmarks for Agentic Swarm Intelligence}

A critical missing component in today's agentic AI ecosystem is a \emph{standardized} benchmark suite for autonomous swarms.  
We propose the development of:

\begin{itemize}
    \item \textbf{Gossip-Aware Multi-Agent Benchmark Suite (GAMABS)}
    \item \textbf{Semantic Propagation Challenge (SPC)}
    \item \textbf{Distributed Orchestration Stress Test (DOST)}
\end{itemize}

Analogous to ImageNet for vision or GLUE for language, such benchmarks would enable reproducible evaluation of decentralized communication substrates.  
Given the increasing role of agentic systems in robotics, infrastructure automation, and open agent ecosystems, the establishment of these benchmark standards is a crucial step for the field.

\section{Experimental Setup}

To evaluate how gossip-based dissemination can enhance coordination in agentic AI systems, we outline an experimental setup that isolates the effects of gossip on state sharing, task allocation, and resilience. This section does not report empirical results. Instead, it establishes a reproducible methodology for testing gossip extensions within multi-agent orchestration frameworks such as MCP- or A2A-based ecosystems.

The goal of the experimental design is threefold:  
(i) to measure how quickly and accurately agents converge on shared state;  
(ii) to assess improvements in decentralized task allocation under uncertainty; and  
(iii) to evaluate robustness under partial failures, delays, and environmental perturbations.  
The experiments use controlled toy environments that reflect real-world industrial automation and disaster-response scenarios.

\subsection{Toy Example 1: Smart Factory Coordination}

We construct a simplified factory with five autonomous agents representing machines on a production line. Each agent has three core attributes: instantaneous workload, queue length, and operational status. We simulate a dynamic environment in which tasks arrive stochastically and machines occasionally fail or slow down.

Two communication models are compared:

\begin{enumerate}
    \item \textbf{Structured-only (Baseline):} Agents communicate solely through direct calls using MCP or A2A-style request--response interactions.
    \item \textbf{Structured + Gossip (Proposed):} Agents retain structured messaging for intentional actions but continuously gossip their partial state to randomly selected peers.
\end{enumerate}

The experiment measures how rapidly the system detects overloads, redistributes tasks, and recovers from failures. In the gossip-augmented model, agents receive redundant state updates that enable them to infer the global load without centralised monitoring. The setup requires only lightweight numerical state vectors, making it feasible to execute thousands of iterations efficiently.

\begin{figure}[!htb]
    \centering
    \includegraphics[width=0.75\linewidth]{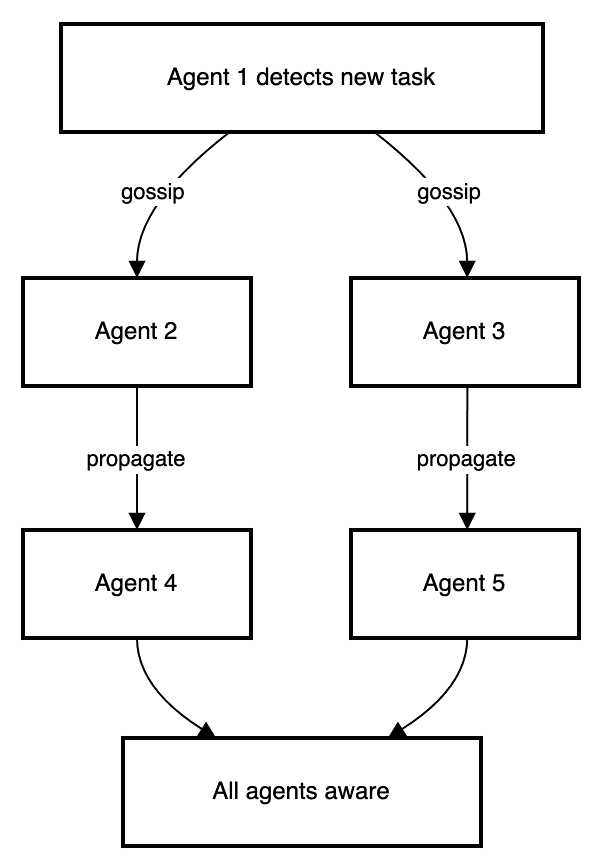}
    \caption{The the experimental flow for Smart Factory Coordination}
    \label{fig:toy-factory}
\end{figure}

A diagrammatic representation of the experimental flow is provided in Fig.~\ref{fig:toy-factory}.

\subsection{Toy Example 2: Disaster-Response Coordination}

The second environment models a post-disaster setting with ten heterogeneous agents (aerial drones and ground robots). The map contains unknown hazards, disconnected regions, and intermittent communication links. Agents must share observations such as survivor locations, blocked paths, and local hazards.

We evaluate two experimental modes:

\begin{enumerate}
    \item \textbf{Direct Messaging Baseline:} Agents share information only with known neighbors through predefined channels.
    \item \textbf{Gossip-Augmented Communication:} Agents opportunistically gossip recent observations whenever communication becomes available, enabling store-and-forward propagation across disconnected regions.
\end{enumerate}

\begin{figure}[!htb]
    \centering
    \includegraphics[width=0.75\linewidth]{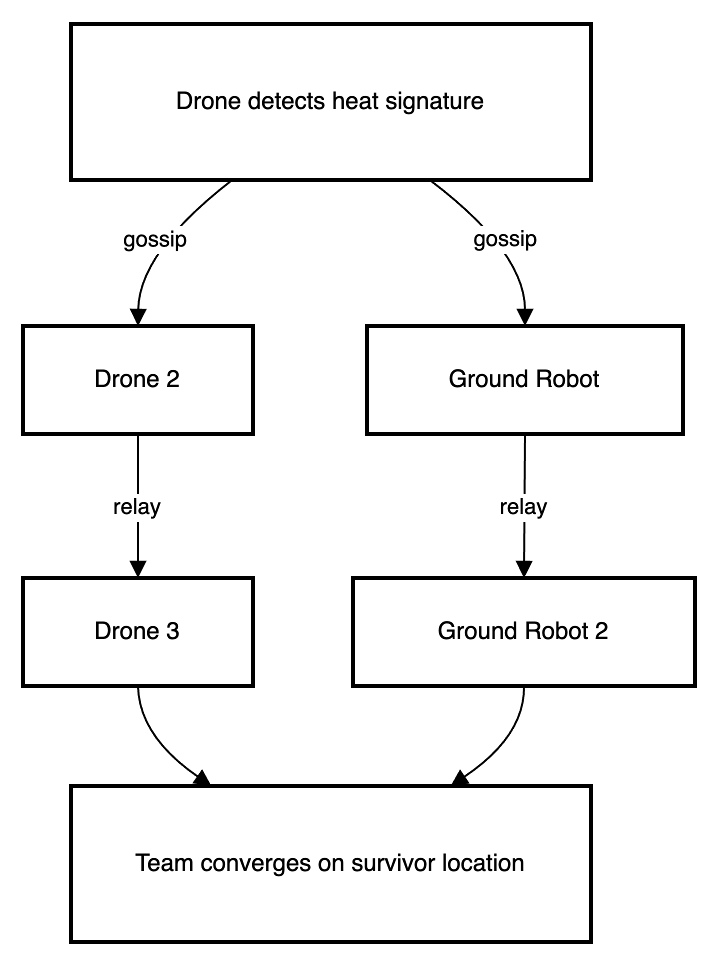}
    \caption{The the experimental flow for Disaster-Response Coordination}
    \label{fig:toy-disaster}
\end{figure}

The experiments simulate link failures, geographic separation, and agent mobility. Metrics include coverage of discovered hazards, propagation delay for critical alerts, and resilience to message loss. This scenario highlights situations where structured communication alone cannot guarantee global situational awareness.the time requiredre~\ref{fig:toy-disaster} illustrates the experimental environment.

\subsection{Evaluation Metrics}

To ensure methodological rigor, both toy environments are evaluated using a shared set of metrics relevant to agentic coordination:

\begin{table}[htb]
\centering
\resizebox{\textwidth}{!}{
\begin{tabular}{lp{0.55\linewidth}}
\toprule
\textbf{Metric} & \textbf{Description} \\
\midrule
State Convergence Time & Time for all agents to reach a consistent global view of shared variables. \\
Staleness Index & Mean age of information available to each agent when a decision is made. \\
Alert Propagation Time & Delay between the first observation of a critical event and system-wide awareness. \\
Task Redistribution Efficiency & Degree to which tasks shift toward underutilized agents after workload changes. \\
Failure Recovery Latency & Time to detect and react to a failed agent. \\
Bandwidth Consumption & Total communication volume generated under each communication protocol. \\
Message Redundancy & Proportion of duplicate or unnecessary message transmissions. \\
Robustness Under Partition & Fraction of agents maintaining coherent coordination despite intermittent partitions. \\
\bottomrule
\end{tabular}
}
\caption{Evaluation metrics for assessing gossip-augmented agentic orchestration.}
\label{tab:evaluation-metrics}
\end{table}

\subsection{Experimental Infrastructure}

All experiments are designed to run in a modular simulation environment implemented in Python. Agents are instantiated as lightweight processes that communicate via a simulated network layer that injects latency, failures, and packet drops. The gossip protocol is implemented following a standard push–pull fan-out design, while structured communication follows the MCP or A2A request--response pattern.

This setup reflects realistic multi-agent deployments while remaining computationally tractable. It allows future research teams to integrate semantic filtering modules, prioritize messages, or embed reinforcement learning policies without altering the core communication substrate.

\subsection{Reproducibility and Extensibility}

The experimental design intentionally avoids assumptions tied to specific agent frameworks. Instead, it provides a neutral environment in which gossip can be layered beneath any structured protocol. This ensures portability and facilitates independent replication.

Future work may extend the setup to include semantic-state gossiping, embedding-based message-relevance filters, or large-scale stress tests across hundreds of agents. Our goal is to establish a systematic starting point for evaluating gossip’s contribution to agentic AI, rather than to claim definitive empirical superiority.

\section{Limitations and Future Work}

Despite the promise of gossip-enhanced coordination for agentic AI, several fundamental limitations remain. These constraints are not merely engineering challenges; they highlight deeper theoretical gaps in our understanding of decentralized intelligence, semantic communication, and large-scale multi-agent behavior. Addressing these issues will be crucial for transforming the Gossip-Enhanced Agentic Coordination Layer (GEACL) from a conceptual substrate into a deployable foundation for future agentic ecosystems.

\subsection{Limitations of the Proposed Framework}

\paragraph{Semantic Dilution and Information Noise.}
GEACL currently treats semantic state as a mergeable object, assuming that CRDT-like reconciliation and periodic exchange are sufficient for meaningful convergence. In practice, agentic systems operate over high-dimensional, non-uniform semantic spaces. As information diffuses stochastically, its semantic fidelity may degrade; frequent merges can blur contextual distinctions or propagate outdated intent. Without stronger models of semantic relevance and decay, gossip risks amplifying noise rather than consolidating knowledge.

\paragraph{Temporal Misalignment and Stale Context.}
Epidemic dissemination introduces inherent delays. While logarithmic convergence is desirable at scale, many agent tasks—especially those involving control, safety, or temporal commitments—demand near-instantaneous coordination. GEACL currently lacks mechanisms for detecting or mitigating the harms of stale information, raising the risk of inconsistent behaviors during rapid environmental changes.

\paragraph{Limited Guarantees for Safety-Critical Coordination.}
Gossip offers probabilistic coverage rather than deterministic delivery. For safety-critical applications—robotic collaboration, healthcare logistics, industrial automation—failure modes cannot rely on eventual consistency alone. GEACL provides no formal guarantees for global ordering, consistency under adversarial conditions, or synchronized commitment. Hybrid deterministic–stochastic coordination remains an open research challenge.

\paragraph{Trust Vulnerabilities in Open Agent Ecosystems.}
GEACL assumes honest, cooperative participation. In open or adversarial environments, malicious agents could inject misleading updates, manipulate merge semantics, or bias peer sampling. Classical gossip is vulnerable to Sybil attacks, collusion, and cascading misinformation. Although cryptographic extensions are possible, a rigorous security model for gossip-based agent communication is still missing.

\paragraph{Scalability to Million-Agent Populations.}
While gossip scales logarithmically in dissemination time, practical scalability depends on bandwidth, merge complexity, and message prioritization. In dense agent ecosystems—urban robotics, industrial fleets, or global digital twins—network pressure, congestion, and merge costs may become prohibitive. GEACL currently lacks hierarchical or locality-preserving mechanisms required for planetary-scale coordination.

\medskip

Taken together, these limitations highlight not only what GEACL enables, but also what it does not yet address. We now turn to the corresponding research avenues that could meaningfully advance the field.

\subsection{Future Research Directions}

\paragraph{Towards Semantically-Aware Gossip.}
Next-generation agent systems require gossip protocols that propagate \emph{meaning}, not merely state. Promising directions include:
\begin{itemize}
    \item embedding-based representations that preserve semantic salience across merges;
    \item ontology-guided diffusion to filter or prioritize domain-relevant concepts;
    \item dynamic propagation policies that modulate fan-out based on relevance, urgency, or predicted global impact.
\end{itemize}
Developing theoretically grounded, semantically enriched gossip remains a major opportunity.

\paragraph{Hybrid Deterministic–Stochastic Architectures.}
GEACL implicitly assumes that gossip and structured protocols coexist harmoniously. However, interaction effects between stochastic background diffusion and deterministic planners remain poorly understood. Key questions include:
\begin{itemize}
    \item how asynchronous gossip influences planner stability and policy convergence;
    \item how agents choose between gossip-derived versus authoritative signals;
    \item how to guarantee safety when contradictory states propagate simultaneously.
\end{itemize}
A mathematical theory of hybrid coordination is likely to be foundational for future agentic infrastructures.

\paragraph{Learning to Gossip: Adaptive, Context-Driven Diffusion.}
Static gossip parameters cannot accommodate the diversity of environments in which agentic systems operate. Learning-based approaches could:
\begin{itemize}
    \item adapt dissemination frequency based on environmental volatility;
    \item select recipients using learned peer-relevance metrics;
    \item choose message content through value-guided communication policies;
    \item co-evolve communication strategies and behavioral policies in MARL settings.
\end{itemize}
This line of work requires new differentiable approximations to epidemic spread.

\paragraph{Secure, Trustworthy Epidemic Communication.}
Robust deployment in open ecosystems depends on addressing trust and provenance. Future work should explore:
\begin{itemize}
    \item cryptographically verifiable gossip with minimal bandwidth overhead;
    \item decentralized provenance tracking capturing causal chains of updates;
    \item anomaly-based detection of malicious propagation behavior;
    \item gossip-resistant defenses against impersonation or collusion.
\end{itemize}
Security must become a first-class component of gossip-enhanced agent systems.

\paragraph{Standardized Benchmarking and Evaluation Protocols.}
Current evaluation methods lack standardization, making comparison across systems difficult. A comprehensive benchmark suite should assess:
\begin{itemize}
    \item semantic convergence accuracy under churn and uncertainty;
    \item robustness against adversarial contamination;
    \item emergent coordination in structured (industrial) and unstructured (disaster-response) settings;
    \item scalability under heterogeneous network and latency conditions.
\end{itemize}
Such benchmarks are essential for reproducible science.

\paragraph{Scaling to Million-Agent Ecosystems.}
Emerging domains—planetary-scale sensor networks, autonomous IoT swarms, global digital twins—require gossip systems capable of managing millions of participants. Promising research avenues include:
\begin{itemize}
    \item hierarchical and cluster-aware epidemic dissemination;
    \item region-constrained and topology-adaptive fan-out strategies;
    \item bandwidth-aware semantic compression;
    \item analytical bounds for ultra-large-scale convergence and stability.
\end{itemize}
These challenges extend classical gossip theory into a new regime.

\medskip

In summary, while GEACL provides a compelling substrate for decentralized coordination, substantial theoretical, algorithmic, and practical work remains. Achieving semantically robust, secure, and scalable gossip-driven communication will be instrumental for enabling agentic AI systems that operate reliably in complex, real-world environments.

\section{Conclusion}
This paper argues that the communication foundations of large-scale, adaptive multi-agent systems require re-examination as agent populations become increasingly heterogeneous, mobile, and context-dependent. Existing frameworks---including MCP, A2A, and related RPC-based interfaces---provide the determinism, reliability, and semantic explicitness needed for tool invocation and structured task execution. Yet, these protocols presuppose stable registries, well-defined request–response cycles, and largely synchronous coordination. As the scale and autonomy of agentic ecosystems grow, such assumptions become limiting: agents must maintain shared situational awareness and coordinate under uncertainty without relying solely on explicit messages or centralised orchestration.

Within this landscape, gossip protocols offer a complementary substrate rather than a substitute for structured communication. The central insight of this work is that gossip functions as a diffuse, continuously evolving informational fabric that aligns local perspectives across a population. Through probabilistic propagation, partial-view sampling, and anti-entropy reconciliation, gossip enables soft forms of coordination that structured protocols cannot easily provide—such as ambient dissemination of context, dynamic peer discovery, rapid failure awareness, and gradual convergence of local world models. By maintaining a background layer of distributed knowledge flow, gossip reduces the burden on deterministic channels, allowing MCP- or A2A-style protocols to be invoked only when precision or transactional guarantees are essential.

This layered view parallels coordination phenomena observed in natural systems, where collective behaviour emerges from the interplay of fast, explicit signalling and slow, ambient cues that diffuse through the environment. In artificial agent ecosystems, gossip plays an analogous role: it supports distributed sensemaking, facilitates adaptive redistribution of tasks, and provides resilience when communication pathways degrade or organisational boundaries shift. Crucially, gossip enables these capabilities without the need for global synchronisation, thereby supporting scalability and robustness in open or adversarial environments.

However, integrating gossip into agentic architectures introduces non-trivial challenges. The stochasticity that enables robustness also complicates semantic interpretation; agents must distinguish informative updates from noise and avoid propagating stale or misleading signals. Furthermore, the redundancy that underpins resilience can amplify adversarial behaviour if authentication, reputation tracking, or multi-source validation are absent. These limitations highlight the need for semantic filtering, trust-aware propagation, and hybrid models that combine gossip with stronger consistency mechanisms at critical decision points.

The industrial automation and disaster-response examples presented in this paper illustrate how gossip operates across distinct operational regimes. In structured environments, gossip enhances fault detection and redundancy; in highly dynamic or fragmented settings, it becomes the primary mechanism for achieving shared awareness. The proposed experimental methodology provides a foundation for evaluating these behaviours systematically, enabling future empirical work to quantify the interaction between gossip-based substrates and structured communication layers.

Looking forward, hybrid architectures that combine decentralised knowledge diffusion with explicit semantic protocols appear particularly promising. Just as modern distributed databases pair eventual-consistency replication with consensus-based coordination for critical operations, future agentic systems may pair gossip-driven situational alignment with strongly structured negotiation and planning. Such an approach could yield multi-agent ecosystems that are simultaneously scalable, adaptive, and semantically coherent.

More broadly, this work suggests a conceptual shift in how communication is understood within agentic AI. Rather than viewing communication solely as a mechanism for executing predefined workflows, it becomes the medium through which collective intelligence emerges. Gossip, with its redundancy, locality, and asymptotic convergence, provides a powerful substrate for this emergence. By integrating gossip with semantically rich, structured protocols, we can begin to design agentic systems capable of robust distributed reasoning, cooperative adaptation, and resilience at scale.

In summary, gossip protocols constitute an underutilised but essential foundation for future multi-agent architectures. Their value lies not in supplanting structured communication frameworks but in augmenting them by filling critical gaps in discovery, resilience, and distributed sensemaking. As agentic AI scales into open, multi-institutional, and safety-critical domains, hybrid communication models that combine the strengths of gossip and structured protocols are likely to define the next generation of autonomous systems. This work provides both the conceptual grounding and the methodological tools necessary to explore these hybrid architectures in future empirical research.

\bibliographystyle{unsrt}
\bibliography{references}

\end{document}